\documentclass[11pt]{article}

\usepackage{graphicx,amssymb,epsfig,amsmath,amssymb,color,cite}

\newtheorem{theorem}{Theorem}
\newtheorem{lemma}{Lemma}

\begin{document}

\title{\textbf{Interaction between cluster synchronization and epidemic spread in community networks}}

\author{Zhongpu Xu$^1$, Kezan Li$^2$, Mengfeng Sun$^1$, Xinchu Fu$^{1,}$\thanks{Corresponding author. Tel: +86-21-66132664; Fax: +86-21-66133292; E-mail: xcfu@shu.edu.cn} \\ \\
$^1${\small Department of Mathematics, Shanghai University, Shanghai 200444, People's Republic of China}\\
$^2${\small School of Mathematics and Computing Science, Guilin University of Electronic Technology,} \\
{\small Guilin 541004, People's Republic of China}}

\date{(\today)}

\maketitle

\begin{abstract}
\noindent In real world, there is a significant relation between human behaviors and epidemic spread. Especially, the reactions among individuals in different communities to epidemics may be different, which lead to cluster synchronization of human behaviors. So, a mathematical model that embeds community structures, behavioral evolution and epidemic transmission is constructed to study the interaction between cluster synchronization and epidemic spread. The epidemic threshold of the model is obtained by using $Ger\breve{s}gorin$ Lemma and dynamical system theory. By applying the Lyapunov stability method, the stability analysis of cluster synchronization and spreading dynamics are presented. Then, some numerical simulations are performed to illustrate and complement our theoretical results. As far as we know, this work is the first one to address the interplay between cluster synchronization and epidemic transmission in community networks, so it may deepen the understanding of the impact of cluster behaviors on infectious disease dynamics.

\medskip

\end{abstract}

\section{Introduction}

 Many natural and artificial systems, such as electricity and transportation systems~\cite{1,2}, biological systems~\cite{3,4}, economic and financial networks~\cite{5,6}, social systems~\cite{7,8}, can be considered as dynamical complex networks. A node in a network denotes its fundamental element, it generally has different meanings in various networks. For instance, human brain can be regarded as a network of neurons linked by synapses, the Internet is made up of routers by physical connections, and so on. A population can also be represented by a complex network in which the nodes represent individuals and the links represent various interactions among individuals, and infectious diseases can only spread along edges of the network. Disease spread on complex networks is an important topic in mathematical epidemiology, and has obtained many research achievements~\cite{9,10,11,12,13a}.

Progress in the research of complex networks are also remarkable, which lead to a deeper study and practical applications to fields related to complex networks. In addition to infectious disease spread, one of the important research topics is synchronization of dynamical networks, which has been widely  studied~\cite{13,14,15,16,17,18}. Synchronization mechanism has been found to be an universal phenomenon in nature and real life, such as a chorus of frogs at summer night, the synchronization of glowworm luminescence, the gradual synchronization of the audience applause frequency in the theater~\cite{19,20,21}, and so on. With the development of synchronization study, there are many widely studied synchronization patterns in literatures, such as adaptive synchronization, cluster synchronization, phase synchronization, and so on.

Synchronization dynamics and epidemic dynamics are two hot topics in network science, however, they are apparently two quite distinct behaviors, so the study of their relationship is very little. Research results show that the correlation between different dynamical behaviors is an interesting direction in order to understand them better in recent years~\cite{22,23,13a}. In an epidemic spreading network, during the spread of the contagion, the disease information also spreads  on the network, so individuals will change their behaviors spontaneously to protect them free from being infected, such as washing hands frequently with soap, avoiding assembly, and so on. This means that the spread of individual awareness during the transmission of an infectious disease can give rise to spontaneously collective risk-averse behaviors among individuals. In Ref.~\cite{22}, the authors have described these consistent behaviors, which result from epidemic spread, as epidemic induced synchronization, then they construct two mathematical models on a complex dynamical network and an epidemic spreading network based on heterogeneous mean-field theory, and study the local and global stabilities of epidemic synchronization. In~\cite{23} Li et al. established two models corresponding to two collective behaviors of dynamical behavior networks, i.e., global synchronization and phase synchronization, then analyzed the control problem. These research results provide a new angle to address the synchronization mechanism of dynamical networks. However, the research is based on a network without community structures.

Moreover, most dynamical models have considered cluster synchronization based on networks with identical nodes in current literatures, which mean the local dynamics of individual nodes are all the same. These indicate current study of synchronization mainly in a group, no synchronization appears between any two different groups. However, it is unreasonable that all the nodes are the same in reality. For example, in social community network, the individual nodes can be regard as the identical function units, the population and the epidemic spreading rate are different in every community, so the correlations between any pair of nodes in different communities are different~\cite{24,25}. Wang et al.~\cite{26} have studied the cluster synchronization of community networks with nonidentical nodes, where the local dynamics of nodes in different communities are different. So, in order to more reasonable, we will combine the collective network and spreading dynamical network, then address interplay between the dynamical networks with community structure and nonidentical nodes and spreading dynamics on complex networks.

Therefore, in this paper, we first establish a model reflecting collective behavior and spreading dynamics on a complex network, where the epidemic spreading model is in a quenched network and dynamical network has community structure and nonidentical nodes, while the local dynamics for all individual nodes are identical in each community, but any pair of nodes in different communities may be different. First, we give the threshold of spreading dynamics network model, then the global stability under certain conditions for this network is considered. By applying the Lyapunov stability method, the stability of global synchronization and spreading dynamics are investigated. Moreover, some numerical simulations are performed to illustrate and complement these theoretical results.

This paper is organized as follows. In Section~2, a dynamical model for community networks with nonidentical nodes and a spreading dynamics network in a quenched network are introduced; then we construct a concrete model of collective behavior and spreading dynamics on complex networks; in Section~3, we estimate the epidemic threshold of spreading dynamics network, and perform stability analysis of global synchronization and spreading dynamics; these theoretical results are illustrated by some numerical simulations given in Section~4; finally,  conclusions are summarized in Section~5.

\section{Models and analysis}

Before consider the community networks with nonidentical nodes, we first make the following assumptions. Suppose that there are $N$ nodes and $m$ communities in these networks, in which $2\le m < N$. Denote by $\sigma_i$  that node $i$ belongs to the $\sigma_i$-th community, $1\le \sigma_i \le m,~~i=1,2,\cdots, N$. The set of all nodes in the $\sigma_i$-th community is denoted by $\mathcal{N}_{\sigma_i}$, and the cardinal number of $\mathcal{N}_{\sigma_i}$ is denoted by $N_{\sigma_i}$. So, if $j\in \mathcal{N}_k, 1\le j\le N, 1\leq k\leq m$, then $\sigma_j=k$. Let $\tilde{\mathcal{N}}_{\sigma_i}$ be the set of all nodes in the $\sigma_i$-th community having direct connections to the nodes in other communities. Obviously, $\tilde{\mathcal{N}}_{\sigma_i}\subseteq \mathcal{N}_{\sigma_i}$.

For simplicity, and in order to more intuitive see each node in the community, we give the following vector to show any node $i$ belongs to its community (This can always be done by renumbering the nodes):

$$(1,2,\cdots,N)=(\overbrace{1,\cdots,N_1}^{first\ community},\overbrace{N_1+1,\cdots,N_1+N_2}^{second\ community},\cdots,\overbrace{N-N_m+1,\cdots,N}^{m-th\ community}),$$
where $N=N_1+\cdots+N_m.$


We present a graphical representation of a complex network with community structure in Fig.~1. We use $f_i$ to denote the local dynamics in the $i$-th community, with each node being an $n$-dimensional dynamical system.

\subsection{Complex dynamical network model}

In the community networks that any node is an n-dimensional dynamical system, and the local dynamics for all nodes in each community are identical, while any pair of nodes in different communities are different. The state equation of the network with time-varying coupling strength can be described by
\begin{equation}\label{e01}
\dot{x}_{i}(t)=f_{\sigma_i}(x_{i}(t))-c_{\sigma_i}(t)\sum\limits_{j=1}^{N}l_{ij}\Gamma x_{j}(t),~~~i=1,2,\cdots,N,
\end{equation}
where $x_{i}(t)=(x_{i1}(t),x_{i2}(t),\cdots,x_{in}(t))^{T}\in \mathbb{R}^{n}$ is the state variable of the node $i$ at time $t\in[0,+\infty)$. The function $f_{\sigma_i}: \mathbb{R}^{n}\rightarrow \mathbb{R}^{n}$ is a continuous nonlinear vector-valued function, which describes the local dynamics for nodes in the $\sigma_i$-th community, which is showing ample dynamical behaviors, such as  equilibrium points, periodic orbits, and chaotic states. The local dynamics for individual nodes in each community are identical, while those of any pair of nodes in different communities are different. Obviously, if $\sigma_i\neq\sigma_j$, i.e., node $i$ and $j$ are geared to the different communities, then $f_{\sigma_i}\neq f_{\sigma_j}$. $c_{\sigma_i}(t)>0$ denotes the time-varying coupling strength in the $\sigma_i$-th community that can be adjusted. $\Gamma=diag(\gamma_{1},\gamma_{2},\cdots,\gamma_{n})\in \mathbb{R}^{n\times n}$ represents the inner-coupling matrix, for simplicity, which is supposed to be a diagonal matrix with $\gamma_i\geq0$. The Laplacian matrix of the whole network $L=(l_{ij})_{N\times N}$ is defined as follows: if node $i$ and node $j~(i\neq j)$ are linked by an edge, then $l_{ij}=l_{ji}=1$; otherwise, $l_{ij}=l_{ji}=0$, and the diagonal elements of the Laplacian matrix $L$ are defined as follows:
\begin{equation}\label{eq01}
l_{ii}=-\sum_{j=1\atop j\neq i}^{N}l_{ij}=-\sum_{j=1\atop j\neq i}^{N}l_{ji}=k_{i},~~~i=1,2,\cdots,N,
\end{equation}
where $k_{i}$ denotes the degree of node $i$.

Moreover, we assume that $L$ is an irreducible matrix, it induces that zero is the smallest eigenvalue of matrix $L$ with single multiplicity and all the other eigenvalues are strictly positive. Hence, by (\ref{eq01}) and matrix theory, there exists a unitary matrix $U$ such that $L=U\Lambda U^{T}$, where $U^{T}U=I$ and $\Lambda={\rm diag}(\lambda_{1},\lambda_{2},\cdots,\lambda_{N})$ with $0=\lambda_{1}<\lambda_{2}\leq\cdots\leq\lambda_{N}$.

Here we suppose that there exists a synchronization orbit $s_{\sigma_i}(t)$ for the system, so we can define the error variables $e_i(t)=x_i(t)-s_{\sigma_i}(t), i=1,\cdots,N$, where $\dot{s}_{\sigma_i}(t)=f_{\sigma_i}(s_{\sigma_i}(t))$, which describes the identical local dynamics for the nodes in the $\sigma_i$-th community.
The dynamical network is said to achieve synchronization if
\begin{equation}\label{e01a}
\lim_{t\rightarrow +\infty}\|e_i(t)\|=0,~~~i=1,\cdots,N,
\end{equation}
which means that the nodes in $\sigma_i$-th community are fully synchronized to dynamic state $s_{\sigma_i}(t)$, while the behavior of nodes in different communities are all independent. Define a set $\Psi=(s_{\sigma_1},s_{\sigma_2},\cdots,s_{\sigma_N})\subset\mathbb{R}^{n\times N}$ as the cluster synchronization manifold for the network~(\ref{e01}). In fact, condition (\ref{e01a}) implies that the synchronization manifold is stable.

\begin{figure}[]
\begin{center}
\includegraphics[width=5cm,height=5cm]{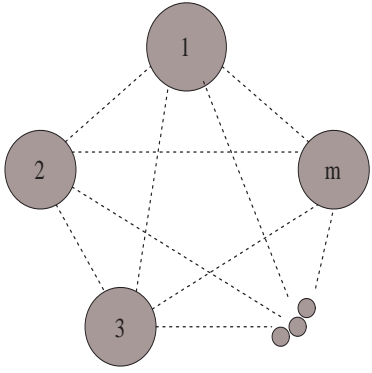}
\caption{The community structure of a complex network. The $i$-th community in this network is denoted by the $i$-th numbered gray circle, with local dynamics $f_i$. The dotted lines show the probable connections
between the $m$ communities.}\label{fig1}
\end{center}
\end{figure}

\subsection{Spreading dynamics network model}

As is well-known, the SIS epidemic model is a classical epidemic spreading model. Hence, we consider that each individual node in the network with community structure can be in one of two distinct states at each time: either susceptible (S) or infected (I). Each infected node becomes susceptible with rate $\delta$, and each susceptible node has a probability $\nu$ of contagion with each infected neighbor. So, we define the effective spreading rate $\lambda=\nu/\delta$. In this paper, we consider the standard SIS model in a quenched network of size $N$, i.e., the community networks are embedded in the quenched network. Let $\rho_{i}(t)$ be the infection density of node $i$ at time $t$. Suppose the relationships between infected and susceptible nodes are uncorrelated, then the evolution equation of node $i$ can be written by
\begin{equation}\label{eq02}
\dot{\rho}_{i}(t)=-\rho_{i}(t)+\lambda\phi_{\sigma_i}(t)[1-\rho_{i}(t)]\sum_{j=1}^{N}a_{ij}\rho_{j}(t),~~~i=1,2,\cdots,N,
\end{equation}
where the infection rate $\lambda\in(0,1]$, and $a_{ij}$ is an element of the adjacency matrix, defining as follows: if there is an edge between nodes $i$ and $j$, then $a_{ij}=1$, or $a_{ij}=0$ otherwise, and the disease prevalence $\rho(t)=\sum\limits_{i=1}^{N}\rho_i(t)/N$. The term $\phi_{\sigma_i}(t)$ denotes the information of synchronization or  the infection control behavior of the individuals, and more details will be given later.

\subsection{Epidemic synchronization model}
In the community networks, information transmission is accompanied by the spread of infectious diseases, which can induce individuals spontaneously reduce the frequency of contact with someone and take the protective measures. That means the rate of the change of the coupling strength $\dot{c}_{\sigma_i}(t)$ in the $\sigma_i$-th community is directly proportional to the infection density $\rho_{\sigma_i}(t)$ of the community. On the other hand, when the collective protective behavior is prevalent, the protective information communication among individuals will become stable because of they have reached a agreement. Thus, the proportional relation between $\dot{c}_{\sigma_i}(t)$ and the synchronization error $\sum\limits_{j\in \mathcal{N}_{\sigma_i}}e_{j}^{T}(t)e_{j}(t)$ is remains valid.

Based on the above assumptions and analysis, combine the traditional models (\ref{e01}) and (\ref{eq02}), we can construct the following SIS epidemic cluster synchronization model:
\begin{equation}\label{e02}
\left\{\begin{array}{ll}
\dot{x}_{i}(t)=f_{\sigma_i}(x_{i}(t))-c_{\sigma_i}(t)\sum\limits_{j=1}^{N}l_{ij}\Gamma x_{j}(t),  \\
\dot{\rho}_{i}(t)=-\rho_{i}(t)+\lambda\phi_{\sigma_i}(t)[1-\rho_{i}(t)]\sum\limits_{j=1}^{N}a_{ij}\rho_{j}(t),~~~i=1,2,\cdots,N, \\
\dot{c}_{\sigma_i}(t)=\beta\rho_{\sigma_i}(t)\sum\limits_{j\in \mathcal{N}_{\sigma_i}}e_{j}^{T}(t)e_{j}(t).
\end{array}\right.
\end{equation}

Based on the above model, we continue the analysis of the epidemic network. Besides, we define the additional term $\phi_{\sigma_i}(t)=(1-\alpha)E_{\sigma_i}(t)+\alpha$ in equation (\ref{eq02}) with constant $\alpha\in(0,1)$, where
$$E_{\sigma_i}(t)=\frac{1}{N_{\sigma_i}}\sum_{j\in \mathcal{N}_{\sigma_i}}\frac{\|s_{\sigma_i}(t)-x_j(t)\|^2}{1+\|s_{\sigma_i}(t)-x_j(t)\|^2}\in[0,1),$$
where $s_{\sigma_i}(t)$ is the synchronous state of the dynamical behavior network in the $\sigma_i$-th community, and $\tilde{E}_{\sigma_i}(t)=\sum\limits_{j\in \mathcal{N}_{\sigma_i}}\|s_{\sigma_i}(t)-x_j(t)\|^2=\sum\limits_{j\in \mathcal{N}_{\sigma_i}}e_{j}^{T}(t)e_{j}(t)$, the parameter $\beta>0$, and $\rho_{\sigma_i}(t)=\sum\limits_{j\in \mathcal{N}_{\sigma_i}}\rho_j(t)/N_{\sigma_i}$ denotes the disease prevalence of the $\sigma_i$-th community. There are $m$ communities in the network, so $E_1(t)=\sum\limits_{j=1}^N\|s_{\sigma_j}(t)-x_j(t)\|^2=\sum\limits_{j=1}^{N}e_{j}^{T}(t)e_{j}(t)=\sum\limits_{j\in\mathcal{N}_1}e_{j}^{T}(t)e_{j}(t)+\cdots+\sum\limits_{j\in\mathcal{N}_m}e_{j}^{T}(t)e_{j}(t)$.

Besides, the term $\phi_{\sigma_i}(t)$ in equation $(\ref{eq02})$ represents the information of synchronization, which is regarded as the admission rate~\cite{27}, therefore, it can be considered as a kind of individual awareness or risk perception. When all individuals in different communities achieve synchronization, i.e., $E_{\sigma_i}(t)\rightarrow0$ as $t\rightarrow\infty$, then $\phi_{\sigma_i}(t)$ is the minimum $\alpha$. If the value of parameter $\alpha$ is smaller, then awareness to collective behavior is greater. The case $\alpha=1$ means that no awareness to the information of synchronization.

The variable $\phi_{\sigma_i}(t)$ also can be quantified as the infection control behavior of all individuals within each community, which means that the behavior of the individual reduces the risk of infection to some extent. So, the infection rate $\nu$ of each susceptible node becomes $\phi_{\sigma_i}(t)\nu$, then the effective infection rate $\lambda$ becomes $\phi_{\sigma_i}(t)\lambda$. Moreover, this individual behavior can be observed by others, for instance, washing hands with sanitizer, avoiding assemblage, wearing face masks, and so on. Collective behavior could embody this individual behavior. In reality, there is a collectivization situation, if more people are wearing face masks in public, which is easier to accept and more people likely to follow these, or vice versa. Hence, the dynamical behavior of this parameter $\phi_{\sigma_i}(t)$ exhibits a synchronization, which could influence the disease dynamics.

\section{Stability analysis of global synchronization and spreading dynamics}

In this section, we will study the global stability of cluster synchronization of model (\ref{e02}) by applying the Lyapunov function method. Predicting the epidemic threshold is an important issue for studying the epidemic spreading network. First, a Lemma is introduced before we address the epidemic threshold of the spreading network.

\begin{lemma}[$Ger\breve{s}gorin$~\cite{28}]\label{lem1} Let $A=(a_{ij})\in C^{n\times n}$ and let $r_i=\sum\limits_{j=1\atop j\neq i}^{n}|a_{ij}|,~~ i=1,2,\cdots,n.$
Then all the eigenvalues of $A$ lie in the union of $n$ closed discs $\cup_{i=1}^n\{z\in\mathbb{C}: |z-a_{ii}|\leq r_i\}.$
\end{lemma}

Now, we give the following Theorem to show the transmission threshold of the epidemic spreading network.

\begin{theorem}\label{thm1}
If an epidemic spreads and becomes endemic, then it is necessarily true that $\lambda>\lambda_c=1/\alpha\rho(A)$, where $\rho(A)$ is the spectral radius of the matrix $A=(a_{ij})_{N\times N},~i,j=1,2,\cdots,N$.
\end{theorem}

\textbf{Proof.} \ From Eq.~(\ref{eq02}) and $\phi_{\sigma_i}(t)\geq\alpha>0$, we obtain
\begin{eqnarray*}
  \dot{\rho}_{i}(t) &=& -\rho_{i}(t)+\lambda\phi_{\sigma_i}(t)[1-\rho_{i}(t)]\sum_{j=1}^{N}a_{ij}\rho_{j}(t) \\
                    &\geq& -\rho_{i}(t)+\lambda\alpha[1-\rho_{i}(t)]\sum_{j=1}^{N}a_{ij}\rho_{j}(t),
\end{eqnarray*}
where $i=1,2,\cdots,N$.

Now, we can consider the equation
\begin{equation}\label{e03}
\dot{\rho}_{i}(t) = -\rho_{i}(t)+\lambda\alpha[1-\rho_{i}(t)]\sum_{j=1}^{N}a_{ij}\rho_{j}(t).
\end{equation}
Then, we calculate the steady-state probability of infection for each node $i$ from (\ref{e03}), denoted by $\rho_{i}$, which is determined by the following nonlinear equation
\begin{equation}\label{e10}
\rho_{i}=\frac{\lambda\alpha\sum_{j}a_{ij}\rho_{j}}{1+\lambda\alpha\sum_{j}a_{ij}\rho_{j}},~~~i=1,2,\cdots,N.
\end{equation}

Letting $\Theta=\sum\limits_{j=1}^{N}a_{ij}\rho_{j}$, then $\rho_{i}=\frac{\lambda\alpha\Theta}{1+\lambda\alpha\Theta}$, then we obtain a self-consistency equation:
\begin{equation}\label{e101}
\Theta=\sum_{j}a_{ij}\frac{\lambda\alpha\Theta}{1+\lambda\alpha\Theta}\equiv f(\Theta).
\end{equation}

Obviously, $\Theta=0$ is always a solution of (\ref{e101}), i.e., $f(0)=0$. Note that
$$f'(\Theta)=\sum_{j}a_{ij}\frac{\lambda\alpha}{(1+\lambda\alpha\Theta)^2}>0,$$
$$f''(\Theta)=\sum_{j}a_{ij}\frac{-2\lambda^2\alpha^2}{(1+\lambda\alpha\Theta)^3}<0,$$
therefore, a nontrivial solution exists only if
\begin{eqnarray}\label{e102}
\nonumber \frac{df(\Theta)}{d\Theta}\mid_{\Theta=0} &=& \sum_{j}a_{ij}\frac{\lambda\alpha}{(1+\lambda\alpha\Theta)^2}|_{\Theta=0}\\
     &=& \sum_{j}\lambda\alpha a_{ij}>1.
\end{eqnarray}
so we have
\begin{equation}\label{e102a}
\lambda>1/\alpha\sum_{j}a_{ij}.
\end{equation}

Since the values of $\sum_{j}a_{ij}$ are related to $i$, so
\begin{equation}\label{e102b}
\lambda> \frac{1}{\alpha\max\limits_{i}\{\sum\limits_{j}a_{ij}\}}.
\end{equation}

From the above inequality, there exists $k\in\{1,2,\cdots,N\}$, such that $\bar{\rho_k}>0$ for the system (\ref{e03}).

According to the matrix $A=(a_{ij})_{N\times N}$, the N eigenvalues of the matrix are $\Lambda_m,\ m=1,2,\cdots,N$, from Lemma $1$, we have
\begin{equation}\label{e102c}
|\Lambda_m|\leq\max_i\{\sum_{j}a_{ij}\},\ m=1,2,\cdots,N.
\end{equation}

So, we obtain
\begin{equation}\label{e102c}
\max_m\{|\Lambda_m|\}\leq\max_i\{\sum_{j}a_{ij}\}.
\end{equation}

Denote by $\rho(A)$ the spectral radius of the matrix $A$, then $\rho(A)=\max_m\{|\Lambda_m|\}$, and
\begin{equation}\label{e102d}
\rho(A)\leq\max_i\{\sum_{j}a_{ij}\}\Rightarrow\frac{1}{\rho(A)}\geq\frac{1}{\max_i\{\sum_{j}a_{ij}\}}.
\end{equation}

If $\lambda>\frac{1}{\alpha\rho(A)}$, there will be $\bar{\rho_k}>0$. So, there exists a $k$, such that $\bar{\rho_k}>0$. Therefore, according to the comparison theorem, we know for the original equation, $\rho_k\geq\bar{\rho_k}>0$. Then, when $\lambda>\frac{1}{\alpha\rho(A)}$, there exists $k\in\{1,2,\cdots,N\}$, such that $\rho_k>0$.

The value of $\lambda$ satisfies the inequality defines the critical epidemic threshold of the original equation,
$$\lambda_c=\frac{1}{\alpha\rho(A)}.$$
The proof is therefore completed. \hfill  $\Box$

Define $$H=\{(\rho_1,\rho_2,\cdots,\rho_N)\in \mathbb{R}^N|0\leq\rho_i\leq1,i=1,2,\cdots,N\},$$
and denote by $H^0$ the interior of $H$, where $\rho_0=(0,0,\cdots,0)\in \mathbb{R}^N$. System (\ref{e02}) is said to be uniformly persistent in $H^0$, if there exists a constant $c\in(0,1)$ such that $\liminf_{t\rightarrow\infty}\rho_i(t)>c$ for all $i$ provided $(\rho_1(0),\rho_2(0),\cdots,\rho_N(0))\in H^0$. Then, we have the following stability result.
\begin{theorem}\label{thm2}
 If $\lambda\leq\frac{1}{\rho(A)}$, then $\rho_0$ is the unique equilibrium of the system (\ref{e02}) and it is globally stable in $H$. If $\lambda>\lambda_c=\frac{1}{\alpha\rho(A)}$, then $\rho_0$ is unstable and the system is uniformly persistent in $H^0$.
\end{theorem}
\textbf{Proof.} \ Consider the Lyapunov function $V(t)=\frac{1}{2}\sum\limits_{i=1}^{N}\rho_i^2(t)$. Calculating the derivative $V(t)$ along the solution of (\ref{e02}) gives
\begin{eqnarray}\label{e103}
\nonumber \frac{dV(t)}{dt} &=& \sum_{i=1}^{N}\rho_i(t)\dot{\rho}_i(t)
= \sum_{i=1}^{N}\rho_i(t)\bigg[-\rho_{i}(t)+\lambda\phi_{\sigma_i}(t)[1-\rho_{i}(t)]\sum_{j=1}^{N}a_{ij}\rho_{j}(t)\bigg] \\
\nonumber  &\leq& \sum_{i=1}^{N}\rho_i(t)\bigg[-\rho_{i}(t)+\lambda\sum_{j=1}^{N}a_{ij}\rho_{j}(t)\bigg] \\
\nonumber  &=& \lambda\sum_{i,j=1}^{N}a_{ij}\rho_i(t)\rho_j(t)-\rho(t)^T\rho(t)
= \lambda\rho(t)^T A\rho(t)-\rho(t)^T\rho(t)\\
&=& \rho(t)^T[\lambda\frac{A+A^T}{2}-E_N]\rho(t),
\end{eqnarray}
where $\rho(t)=(\rho_1(t),\rho_2(t),\cdots,\rho_N(t))^T$ and the matrix $A=(a_{ij})_{N\times N}$. From Eq.~(\ref{e103}), we can get
\begin{eqnarray}\label{e103a}
\nonumber \lambda_{\max}[\lambda\frac{A+A^T}{2}-E_N] &=& \lambda_{\max}(\lambda\frac{A+A^T}{2})-1 \\
\nonumber  &\leq&  \lambda\lambda_{\max}(A)-1,
\end{eqnarray}
we know $\rho(A)=\lambda_{\max}(A)$, so $\lambda\leq\frac{1}{\rho(A)}$ leads to $\frac{dV(t)}{dt}\leq0$. In addition, from (\ref{e103}) we get $\dot{V}(t)=0$ if and only if $\rho(t)=0$. Therefore, when $\lambda\leq\frac{1}{\rho(A)}$, the singleton $\{\rho_0\}$ is the only compact invariant subset of the set $\{\rho|\frac{dV(t)}{dt}=0\}$. By LaSalle's Invariance Principle, $\rho_0$ is globally asymptotically stable.

From Theorem $1$, if $\lambda>\frac{1}{\alpha\rho(A)}$, we can conclude that $\frac{dV(t)}{dt}>0$ in a neighborhood of $\rho_0$ in $H^0$. So, the equilibrium $\rho_0$ is unstable, and the instability means the network is uniformly persistent in $H^0$ by a similar discussion in Ref.~\cite{29}.
\hfill  $\Box$ \\

The synchronous state in a network can be defined as $s_{\sigma_i}(t)=\frac{1}{N_{\sigma_i}}\sum\limits_{j\in \mathcal{N}_{\sigma_i}}x_j(t)$. According to the above definition $e_i(t)=x_i(t)-s_{\sigma_i}(t)$ and Eq.~(\ref{e01}), it is easy to obtain $\sum_{i=1}^{N}e_i(t)=0$, we can write the corresponding error system as
\begin{equation}\label{e12}
 \dot{e}_i(t) = f_{\sigma_i}(x_i)-f_{\sigma_i}(s_{\sigma_i})-c_{\sigma_i}(t)\sum\limits_{j=1}^{N}l_{ij}\Gamma e_{j},
\end{equation}
where $i=1,2,\cdots,N$.

By letting $F(t)=(f_{\sigma_1}(x_1(t))^T-f_{\sigma_1}(s_{\sigma_1}(t))^T,\cdots,f_{\sigma_N}(x_N(t))^T-f_{\sigma_N}(s_{\sigma_N}(t))^T)^T$, $C(t)=(c_{\sigma_1}(t),\cdots,c_{\sigma_N}(t))^T$, and $e(t)=(e_1^T,\cdots,e_N^T)^T$, the system (\ref{e12}) can be rewritten as
\begin{equation}\label{e14}
\dot{e}(t)=F(t)-C(t)(L\otimes\Gamma)e(t),
\end{equation}
where $\otimes$ is the Kronecker product.

\textbf{Assumption 1}\
\emph{Suppose that there exists a positive definite diagonal matrix $P=diag(p_{1},p_{2},\cdots,p_{n})$ and a constant $\xi>0$, such that the nonlinear vector-valued continuous function $f_{\sigma_i}(x_i(t))$ satisfies
\begin{equation}\label{e13}
(x(t)-s_i(t))^{T}P[f_i(x(t))-f_i(s_i(t))]\leq\xi(x(t)-s_i(t))^{T}(x(t)-s_i(t)),
\end{equation}
for all $x(t)\in R^{n}$ and $t\geq 0,\ i=1,2,\cdots,m.$ }\\

Now, we have the following Theorem about the global stability for the synchronization manifold $\Psi$ of system (\ref{e02}).
\begin{theorem}\label{thm3}
Suppose that $\lambda_c=1/\alpha\rho(A)>0$ is the epidemic threshold of system (\ref{e02}). If the effective rate $\lambda>\lambda_c$, then there exists a unique endemic equilibrium $\rho^*$ of the model and it is globally asymptotically stable. And the synchronization manifold of the system is also globally asymptotically stable.
\end{theorem}
\textbf{Proof.} \ Since $S_i(t)+\rho_i(t)=1$ for $i=1,2,\cdots,N$, and we set $\beta_{ij}=\lambda a_{ij}$, then the epidemic network can be rewritten as
\begin{eqnarray}\label{e2}
  \dot{S}_i(t) &=& 1-S_i-\sum_j\phi_{\sigma_i}(t)\beta_{ij}S_i\rho_j, \\
  \dot{\rho}_i(t) &=& -\rho_i+\sum_j\phi_{\sigma_i}(t)\beta_{ij}S_i\rho_j.\label{e222}
\end{eqnarray}

If $\lambda>\lambda_c$, then the system has an endemic equilibrium $\rho^*$. Denote the endemic equilibrium by $(\rho_1^*,\rho_2^*,\cdots,\rho_N^*)\in H^0$, $\rho_i^*>0$ for $i=1,2,\cdots,N$.
Then, we set $\bar{\beta}_{ij}=\beta_{ij}S^*_i\rho^*_j$, and $S^*_i(t)=1-\rho^*_i(t)$, and define a matrix

 \begin{equation}\label{e3}     
\bar{B}=\left( 
  \begin{array}{cccc}   
    \sum_{l\neq1}\bar{\beta}_{1l} & -\bar{\beta}_{21} & \cdots & -\bar{\beta}_{N1}\\  
    -\bar{\beta}_{12} & \sum_{l\neq2}\bar{\beta}_{2l} & \cdots & -\bar{\beta}_{N2} \\  
      \vdots & \vdots & \ddots & \vdots \\  
     -\bar{\beta}_{1N} & -\bar{\beta}_{2N} & \cdots &\sum_{l\neq N}\bar{\beta}_{Nl}\\  
  \end{array}
\right),            
\end{equation}
where the sum of each column in the matrix equals zero.

In order to testify the Theorem, we consider a Lyapunov function $V_0(t)=V_1(t)+V_2(t)$, in which

\begin{equation}\label{e4}
V_{1}(t)=\sum_{i=1}^{N}v_i(S_i-S^{\ast}_{i}\ln S_{i}+\rho_{i}-\rho^{\ast}_{i}\ln \rho_i),
\end{equation}
where $v_i>0$ denotes the cofactor of the $i$-th diagonal entry of $\bar{B}$ satisfying $\bar{B}v=0$, $v=(v_1,\cdots,v_N)^T$.

Since a unique endemic equilibrium exists, from Eqs.~(\ref{e2}) and $(\ref{e222})$, we can get
\begin{eqnarray}\label{e5}
1 &=& S_i^*+\alpha\sum_j\beta_{ij}S_i^*\rho_j^*, \\
\rho_i^* &=& \alpha\sum_j\beta_{ij}S_i^*\rho_j^*.\label{e50}
\end{eqnarray}

Then, by using Eqs.~(\ref{e5})-(\ref{e50}), the derivative of $V_1(t)$ along the trajectory of system admits
\begin{eqnarray}\label{e6}
\nonumber \dot{V}_1(t) &=& \sum_{i=1}^{N}v_i\bigg(1-S_i-\sum_j\phi_{\sigma_i}(t)\beta_{ij}S_i\rho_j-\frac{S_i^*}{S_i}\big(1-S_i-\sum_j\phi_{\sigma_i}(t)\beta_{ij}S_i\rho_j\big)-\rho_i \\
\nonumber && +\sum_j\phi_{\sigma_i}(t)\beta_{ij}S_i\rho_j-\frac{\rho_i^*}{\rho_i}\big(\sum_j\phi_{\sigma_i}(t)\beta_{ij}S_i\rho_j-\rho_i\big)\bigg)\\
\nonumber &=& \sum_{i=1}^{N}v_i\bigg(S_i^*+\alpha\sum_j\beta_{ij}S_i^*\rho_j^*-S_i-\sum_j\phi_{\sigma_i}(t)\beta_{ij}S_i\rho_j-\frac{S_i^*}{S_i}\big(\alpha\sum_j\beta_{ij}S_i^*\rho_j^* \\
\nonumber && +S_i^*\big)+S_i^*+\sum_j\phi_{\sigma_i}(t)\beta_{ij}S_i^*\rho_j-\rho_i+\sum_j\phi_{\sigma_i}(t)\beta_{ij}S_i\rho_j \\
\nonumber && -\sum_j\phi_{\sigma_i}(t)\beta_{ij}S_i\rho_j\frac{\rho_i^*}{\rho_i}+\alpha\sum_j\beta_{ij}S_i^*\rho_j^*\bigg)\\
\nonumber &=& \sum_{i=1}^{N}v_i\bigg(-S_i^*\big(\frac{S_i}{S_i^*}+\frac{S_i^*}{S_i}-2\big)+2\alpha\sum_j\beta_{ij}S_i^*\rho_j^*-\alpha\sum_j\beta_{ij}\frac{(S_i^*)^2}{S_i}\rho_j^* \\
&& -\sum_j\phi_{\sigma_i}(t)\beta_{ij}S_i\rho_j\frac{\rho_i^*}{\rho_i}+\sum_j\phi_{\sigma_i}(t)\beta_{ij}S_i^*\rho_j-\rho_i\bigg).
\end{eqnarray}

Since $\frac{S_i}{S_i^*}+\frac{S_i^*}{S_i}-2\geq0$, we get
\begin{equation}\label{e7}
-S_i^*(\frac{S_i}{S_i^*}+\frac{S_i^*}{S_i}-2)\leq0,
\end{equation}
and the equality sign holds if and only if $S_i=S_i^*$.

Now, we will prove that
\begin{equation}\label{e8}
\sum_{i=1}^{N}v_i\bigg(\sum_{j=1}^N\phi_{\sigma_i}(t)\beta_{ij}S_i^*\rho_j-\rho_i\bigg)\leq\frac{E_1(t)}{N_0}\sum_{i,j=1}^{N}v_i\beta_{ij}S_i^*\rho_j.
\end{equation}

For the left-hand side of above inequality, we have
\begin{eqnarray*}\label{e9}
\sum_{i=1}^{N}v_i\bigg(\sum_{j=1}^N\phi_{\sigma_i}(t)\beta_{ij}S_i^*\rho_j-\rho_i\bigg) &=& \sum_{i=1}^{N}\bigg(\sum_{j=1}^N\phi_{\sigma_i}(t)\beta_{ji}S_j^*v_j-v_i\bigg)\rho_i \\
  &\leq&\sum_{i,j=1}^N E_{\sigma_i}(t)v_j\beta_{ji}S_j^*\rho_i+\sum_{i=1}^{N}\bigg(\alpha\sum_{j=1}^N\beta_{ji}S_j^*v_j-v_i\bigg)\rho_i\\
  &\leq& \sum_{i,j=1}^N\frac{\tilde{E}_{\sigma_i}(t)}{N_{\sigma_i}}v_j\beta_{ji}S_j^*\rho_i+\sum_{i=1}^{N}\bigg(\alpha\sum_{j=1}^N\beta_{ji}S_j^*v_j-v_i\bigg)\rho_i \\
&\leq&  \frac{E_1(t)}{N_0}\sum_{i,j=1}^Nv_j\beta_{ji}S_j^*\rho_i+\sum_{i=1}^{N}\bigg(\alpha\sum_{j=1}^N\beta_{ji}S_j^*v_j-v_i\bigg)\rho_i,
\end{eqnarray*}
where $N_0=\min\{N_{\sigma_1},N_{\sigma_2},\cdots,N_{\sigma_N}\}$.

By using $\bar{B}v=0$ we get
\begin{equation*}\label{e10}
   \alpha\sum_{j=1}^N\beta_{ji}S_j^*\rho_i^*v_j = \alpha\sum_{j\neq i}^N\bar{\beta}_{ji}v_j+\alpha\bar{\beta}_{ii}v_i
    = \alpha\sum_{j\neq i}^N\bar{\beta}_{ij}v_i+\alpha\bar{\beta}_{ii}v_i
    = \alpha\sum_{j=1}^N\bar{\beta}_{ij}v_i
    = \rho_i^*v_i,
\end{equation*}
that deduces $\alpha\sum \limits_{j=1}^N\beta_{ji}S_j^*v_j-v_i=0$ for all $i$.

Then, we get inequality (\ref{e8}).

Using Eqs.~(\ref{e6})-(\ref{e8}), and applying $\phi_{\sigma_i}(t)\geq\alpha>0$ and $\bar{\beta}_{ij}=\beta_{ij}S_i^*\rho_j^*$, we further get
\begin{eqnarray}\label{e11}
\nonumber \dot{V}_1(t) &\leq& \frac{E_1(t)}{N_0}\sum_{i,j=1}^{N}v_i\beta_{ij}S_i^*\rho_j \\
\nonumber && +\alpha\sum_{i=1}^Nv_i\bigg(2\sum_{j=1}^N\bar{\beta}_{ij}-\sum_{j=1}^N\bar{\beta}_{ij}\frac{S_i^*}{S_i}-\sum_{j=1}^N\bar{\beta}_{ij}\frac{\rho_jS_i\rho_i^*}{\rho_iS_i^*\rho_j^*}\bigg) \\
\nonumber &=&  \frac{\sum\limits_{i,j=1}^{N}v_i\beta_{ij}S_i^*\rho_j}{N_0}\sum\limits_{i=1}^{N}e_{i}^{T}(t)e_{i}(t) \\
   && +\alpha\sum_{i,j=1}^Nv_i\bar{\beta}_{ij}\bigg(2-\frac{S_i^*}{S_i}-\frac{\rho_jS_i\rho_i^*}{\rho_iS_i^*\rho_j^*}\bigg).
\end{eqnarray}

Based on graph theory, the authors in Ref.~\cite{29} have proven that $$\sum\limits_{i,j=1}^Nv_i\bar{\beta}_{ij}\bigg(2-\frac{S_i^*}{S_i}-\frac{\rho_jS_i\rho_i^*}{\rho_iS_i^*\rho_j^*}\bigg)\leq0$$ for positive $\beta_{ij}$. And
\begin{equation}\label{e141}
V_2(t)=\frac{1}{2}e^{T}(t)(I_{N}\otimes P)e(t)+\frac{1}{2\beta}\bar{\beta}[C_0-C(t)]^T[C_0-C(t)],
\end{equation}
where $C_0=(c_0,\cdots,c_0)^T$ is $N$-dimensional and $c_0$ is a undetermined constant, $\bar{\beta}=\lambda_2\lambda_{\min}(P\Gamma)/N>0$, and $\lambda_{\min}(P\Gamma)$ denotes the minimal eigenvalue of matrix $P\Gamma$. Therefore,
\begin{eqnarray}\label{e15}
\nonumber \dot{V}_2(t) &=& e^{T}(t)(I_{N}\otimes P)[F(t)-C(t)(L\otimes\Gamma)e(t)]-\frac{\bar{\beta}}{\beta}[C_0-C(t)]^T\dot{C}(t)  \\
\nonumber   &=& e^{T}(t)(I_{N}\otimes P)F(t)-C(t)e^{T}(t)(L\otimes P\Gamma)e(t)-\frac{\bar{\beta}}{\beta}\sum_{i=1}^{N}(c_0-c_{\sigma_i}(t))\dot{c}_{\sigma_i}(t) \\
\nonumber   &=& \sum_{i=1}^{N}e_{i}^{T}(t)P[f_{\sigma_i}(x_{i}(t))-f_{\sigma_i}(s_{\sigma_i}(t))]-C(t)e^{T}(t)(L\otimes P\Gamma)e(t) \\
\nonumber    &&  +\bar{\beta}\sum_{i=1}^{N}\bigg(c_{\sigma_i}(t)\rho_{\sigma_i}(t)\sum_{j\in \mathcal{N}_{\sigma_i}}e_{j}^{T}(t)e_{j}(t)\bigg)-\bar{\beta}c_0\sum_{i=1}^{N}\bigg(\rho_{\sigma_i}(t)\sum_{j\in \mathcal{N}_{\sigma_i}}e_{j}^{T}(t)e_{j}(t)\bigg)\\
\nonumber     &\leq&  \xi\sum_{i=1}^{N}e_{i}^{T}(t)e_{i}(t)-C(t)e^{T}(t)(L\otimes P\Gamma)e(t)+\bar{\beta}\sum_{i=1}^{N}\bigg(c_{\sigma_i}(t)\sum_{j\in \mathcal{N}_{\sigma_i}}e_{j}^{T}(t)e_{j}(t)\bigg)\\
              && -\bar{\beta}c_0\sum_{i=1}^{N}\bigg(\rho_{\sigma_i}(t)\sum_{j\in \mathcal{N}_{\sigma_i}}e_{j}^{T}(t)e_{j}(t)\bigg).
\end{eqnarray}

Now, introducing a transformation $y(t)=[y_1^T(t),\cdots,y_N^T(t)]^T=(U^T\otimes I_n)e(t)$, we have
$$\sum_{i=1}^{N}e_{i}^{T}(t)e_{i}(t)=\sum_{i=1}^{N}y_{i}^{T}(t)y_{i}(t).$$
Then we have
\begin{eqnarray}\label{e16}
\nonumber  &&  \xi\sum_{i=1}^{N}e_{i}^{T}(t)e_{i}(t)-C(t)e^{T}(t)(L\otimes P\Gamma)e(t)+\bar{\beta}\sum_{i=1}^{N}\bigg(c_{\sigma_i}(t)\sum_{j\in \mathcal{N}_{\sigma_i}}e_{j}^{T}(t)e_{j}(t)\bigg)\\
\nonumber  && -\bar{\beta}c_0\sum_{i=1}^{N}\bigg(\rho_{\sigma_i}(t)\sum_{j\in \mathcal{N}_{\sigma_i}}e_{j}^{T}(t)e_{j}(t)\bigg)\\
\nonumber  &=&  \xi\sum_{i=1}^{N}e_{i}^{T}(t)e_{i}(t)-C(t)y^{T}(t)(\Lambda\otimes P\Gamma)y(t)+\bar{\beta}\sum_{i=1}^{N}\bigg(c_{\sigma_i}(t)\sum_{j\in \mathcal{N}_{\sigma_i}}e_{j}^{T}(t)e_{j}(t)\bigg)\\
\nonumber  && -\bar{\beta}c_0\sum_{i=1}^{N}\bigg(\rho_{\sigma_i}(t)\sum_{j\in \mathcal{N}_{\sigma_i}}e_{j}^{T}(t)e_{j}(t)\bigg)\\
\nonumber  &=&  \xi\sum_{i=1}^{N}e_{i}^{T}(t)e_{i}(t)-\sum_{i=1}^{N}c_{\sigma_i}(t)\lambda_{i}y_{i}^{T}(t)P\Gamma y_{i}(t)+\bar{\beta}\sum_{i=1}^{N}\bigg(c_{\sigma_i}(t)\sum_{j\in \mathcal{N}_{\sigma_i}}e_{j}^{T}(t)e_{j}(t)\bigg)\\
\nonumber  && -\bar{\beta}c_0\sum_{i=1}^{N}\bigg(\rho_{\sigma_i}(t)\sum_{j\in \mathcal{N}_{\sigma_i}}e_{j}^{T}(t)e_{j}(t)\bigg)\\
\nonumber  &\leq&   \xi\sum_{i=1}^{N}e_{i}^{T}(t)e_{i}(t)-\lambda_{2}\lambda_{\min}(P\Gamma)\sum_{i=1}^{N}c_{\sigma_i}(t)y_{i}^{T}(t)y_{i}(t)+\bar{\beta}N\sum_{i=1}^{N}c_{\sigma_i}(t)e_{i}^{T}(t)e_{i}(t)\\
\nonumber  && -\bar{\beta}c_0\sum_{i=1}^{N}\bigg(\rho_{\sigma_i}(t)\sum_{j\in \mathcal{N}_{\sigma_i}}e_{j}^{T}(t)e_{j}(t)\bigg)\\
\nonumber  &=&  \xi\sum_{i=1}^{N}e_{i}^{T}(t)e_{i}(t)-\lambda_{2}\lambda_{\min}(P\Gamma)\sum_{i=1}^{N}c_{\sigma_i}(t)e_{i}^{T}(t)e_{i}(t)+\bar{\beta}N\sum_{i=1}^{N}c_{\sigma_i}(t)e_{i}^{T}(t)e_{i}(t)\\
\nonumber  && -\bar{\beta}c_0\sum_{i=1}^{N}\bigg(\rho_{\sigma_i}(t)\sum_{j\in \mathcal{N}_{\sigma_i}}e_{j}^{T}(t)e_{j}(t)\bigg)\\
\nonumber  &=&  \xi\sum_{i=1}^{N}e_{i}^{T}(t)e_{i}(t)+[\bar{\beta}N-\lambda_{2}\lambda_{\min}(P\Gamma)]\sum_{i=1}^{N}c_{\sigma_i}(t)e_{i}^{T}(t)e_{i}(t)\\
\nonumber  && -\bar{\beta} c_0\sum_{i=1}^{N}\bigg(\rho_{\sigma_i}(t)\sum_{j\in \mathcal{N}_{\sigma_i}}e_{j}^{T}(t)e_{j}(t)\bigg)\\
\nonumber  &\leq&  \xi\sum_{i=1}^{N}e_{i}^{T}(t)e_{i}(t)-\bar{\beta}c_0\rho_1(t)\sum_{i=1}^{N}\sum_{j\in \mathcal{N}_{\sigma_i}}e_{j}^{T}(t)e_{j}(t) \\
\nonumber  &\leq&   \xi\sum_{i=1}^{N}e_{i}^{T}(t)e_{i}(t)-\bar{\beta}c_0\rho_1(t)\sum_{i=1}^{N}e_{i}^{T}(t)e_{i}(t) \\
\nonumber  &=&   [\xi-\bar{\beta}c_0\rho_1(t)]\sum_{i=1}^{N}e_{i}^{T}(t)e_{i}(t),
\end{eqnarray}
where $\rho_1(t)=\min\{\rho_{\sigma_1}(t),\rho_{\sigma_2}(t),\cdots,\rho_{\sigma_N}(t)\}$.

If the effective rate $\lambda>\lambda_c$ for system (\ref{e02}), there exists $\rho^*\in(0,1]$, such that $\lim_{t\rightarrow+\infty}\rho_1(t)=\rho^*$. By choosing $\varepsilon\in(0,\rho^*)$ and $t_0>0$, so $\rho_1(t)>\rho^*-\varepsilon$ for all $t>t_0$. When $t>t_0$, we have
\begin{equation}\label{e17}
\dot{V}_2(t)\leq[\xi-\bar{\beta}c_0(\rho^*-\varepsilon)]\sum_{i=1}^{N}e_{i}^{T}(t)e_{i}(t).
\end{equation}

Thus, we also can select an adequately large constant $c_0$ such that $\dot{V}_2(t)\leq0$. Furthermore, we can get the following inequalities
$$\frac{\lambda_{\min}(P)}{2}\sum_{i=1}^{N}e_{i}^{T}(t)e_{i}(t)+\frac{\bar{\beta}}{2\beta}\sum_{i=1}^{N}(c_0-c_{\sigma_i}(t))^{2}\leq V_2(t)\leq $$ $$\frac{\lambda_{\max}(P)}{2}\sum_{i=1}^{N}e_{i}^{T}(t)e_{i}(t)+\frac{\bar{\beta}}{2\beta}\sum_{i=1}^{N}(c_0-c_{\sigma_i}(t))^{2}.$$

So, the Lyapunov function $V_2(t)$ has infinitesimal upper bound and is infinitely large.

According to the above discussions, we have
\begin{equation}\label{e17}
\dot{V_0}(t)\leq\bigg[\frac{\sum\limits_{i,j=1}^{N}v_i\beta_{ij}S_i^*\rho_j}{N_0}+\xi-\frac{\lambda_{2}\lambda_{\min}(P\Gamma)}{N}c_0(\rho^*-\varepsilon) \bigg]\sum\limits_{i=1}^{N}e_{i}^{T}(t)e_{i}(t).
\end{equation}

By selecting an adequately large constant $c_0$, we have $\dot{V_0}(t)\leq 0$. By the LaSalle-Yoshizawa theorem, we conclude that the equilibrium of system~(\ref{e02}) at $e_i(t)=0$ and $c_0-c_{\sigma_i}(t)=0$ are globally uniformly stable, that is, the synchronisation manifold $\Psi$ of the network is also globally asymptotically stable.
\hfill  $\Box$

\section{Numerical simulations}
In this section, to illustrate the above theoretical results about the model of SIS epidemic synchronization network (\ref{e02}), some numerical examples are presented. We consider a network with three communities, and the network is Barab$\acute{a}$si-Albert(BA) scale-free network~\cite{30} as the topological structures embedded in model (\ref{e02}). So, the network is produced by integrating three BA networks(three communities) with a few edges connected them randomly. The BA network is produced with initial nodes $m_0=4$ and adding a new node with $m=3$ edges in each time step, the size is set as $N=100$. We set the sizes of the three communities are $N_1=20,N_2=30,N_3=50$, respectively, and they have the same BA network  structures. The local dynamics $f$ of all nodes in three communities are defined as the chaotic Lorenz system, chaotic Chen system, and chaotic L$\ddot{u}$ system, respectively. And their vector field functions are denoted by $f_1, f_2,$ and $f_3$. The chaotic Lorenz oscillator, as the desired orbit, can be described by the system
\begin{equation}\label{e18}
\left\{\begin{array}{ll}
\dot{x}_{1}=a(x_{2}-x_{1}),  \\
\dot{x}_{2}=bx_{1}-x_{2}-x_{1}x_{3},\\
\dot{x}_{3}=x_{1}x_{2}-cx_{3},
\end{array}\right.
\end{equation}
where parameters $a=10, b=28, c=8/3$.

The Chen system is
\begin{equation}\label{e19}
\left\{\begin{array}{ll}
\dot{y}_{1}=d(y_{2}-y_{1}),  \\
\dot{y}_{2}=ey_{1}-y_{1}y_{3}+fy_{2},\\
\dot{y}_{3}=y_{1}y_{2}-gy_{3},
\end{array}\right.
\end{equation}
where parameters $d=35, e=-7, f=28, g=3$.

The L$\ddot{u}$ system is
\begin{equation}\label{e20}
\left\{\begin{array}{ll}
\dot{z}_{1}=h(z_{2}-z_{1}),  \\
\dot{z}_{2}=-z_{1}z_{3}+kz_{2},\\
\dot{z}_{3}=z_{1}z_{2}-lz_{3},
\end{array}\right.
\end{equation}
where parameters $h=36, k=20, l=3$.

In the following simulations, we suppose the inner-coupling matrix $\Gamma=I_{N}$, the initial coupling strength $c(0)=0.001$, $\beta=0.01$, the initial infection density $\rho_3(0)=0.003, \rho_k(0)=0, k=4,\cdots,d_m$, where $d_m$ is the maximum degree. Define the synchronization errors of the three communities as $E_1^*(t)=\sum\limits_{i=2}^{N_1}[x_{1}(t)-x_{i}(t)]^{2}, E_2^*(t)=\sum\limits_{i=N_1+2}^{N_1+N_2}[x_{N_1+1}(t)-x_{i}(t)]^{2},$ and $E_3^*(t)=\sum\limits_{i=N_1+N_2+2}^{N}[x_{N_1+N_2+1}(t)-x_{i}(t)]^{2}$. In this case, by the simulations we can obtain the total number of links in the network is $294$, and the three BA networks with $24$ edges connected them randomly.

\begin{figure}[]
\begin{center}
\includegraphics[width=14cm,height=8cm]{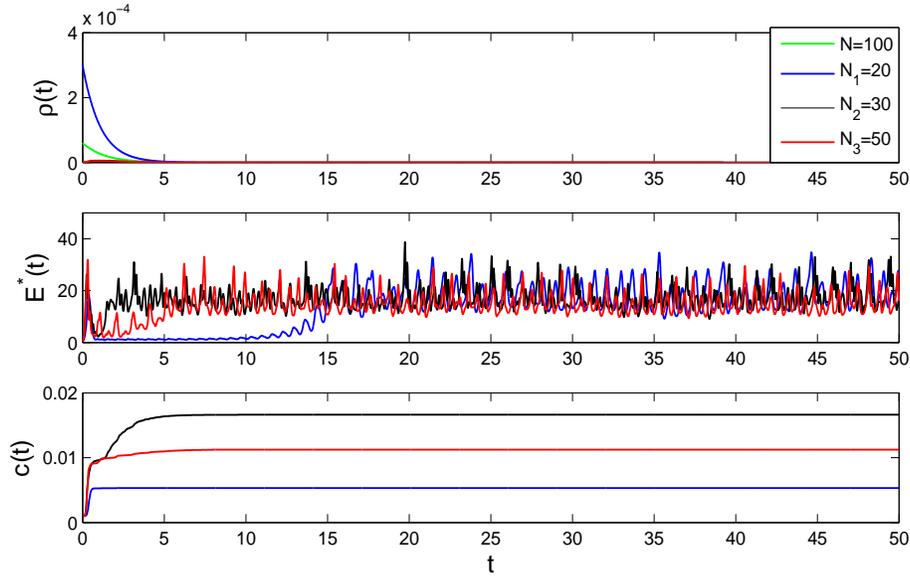}
\caption{The changes of epidemic prevalence $\rho(t)$, synchronization error $E^*(t)$, and coupling strength $c(t)$ in model (\ref{e02}) with epidemic rate $\lambda=0.01$. The network is composed of three randomly connected BA scale-free networks with size $N_1=20, N_2=30, N_3=50$. The only green line denotes the epidemic prevalence of the total network in first subgraph, and $\rho(t), E^*(t),c(t)$ of three communities described by blue, black, red lines in three subgraphs.}\label{fig2}
\end{center}
\end{figure}

We set $\alpha=0.5$. From Fig.~\ref{fig2}, as the growth of time $t$, we get the epidemic prevalence $\rho(t)$ of the total network and three different communities rapidly converge to zero when the effective spreading rate $\lambda=0.01<\frac{1}{\rho(A)}$, i.e., the disease will die out if $\lambda<\frac{1}{\rho(A)}$. In this case, the synchronization errors $E^*(t)$ in three communities do not coverage to zero, which means that the epidemic dynamics cannot successfully induce the synchronization of individuals under small epidemic rate $\lambda=0.01$. In addition, all the time-varying coupling strengths $c(t)$ of three communities reach the steady-state values.

\begin{figure}[]
\begin{center}
\includegraphics[width=14cm,height=8cm]{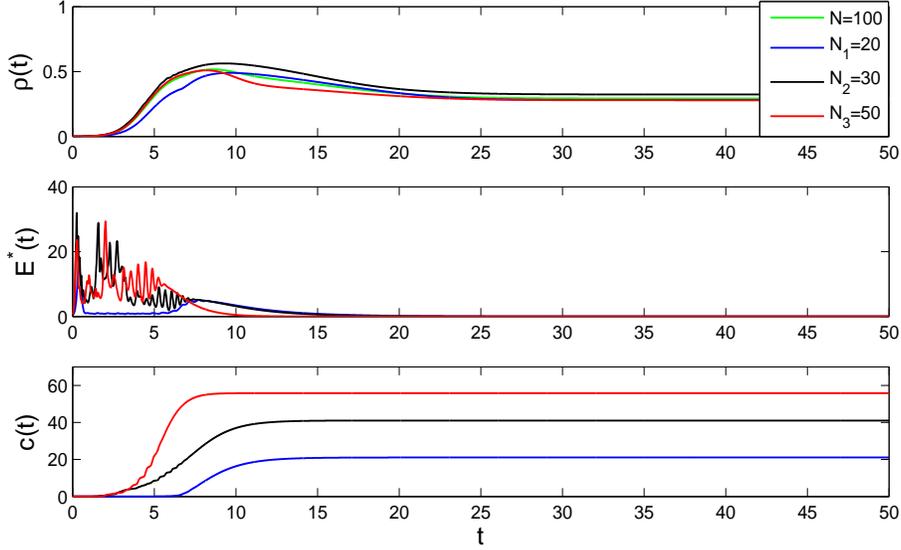}
\caption{The changes of epidemic prevalence $\rho(t)$, synchronization error $E^*(t)$, and coupling strength $c(t)$ in model (\ref{e02}) with epidemic rate $\lambda=0.4$. The network is composed of three randomly connected BA scale-free networks with size $N_1=20, N_2=30, N_3=50$. Line colors are similar to that shown in Figure 2.}\label{fig3}
\end{center}
\end{figure}

\begin{figure}[]
\begin{center}
\includegraphics[width=14cm,height=8cm]{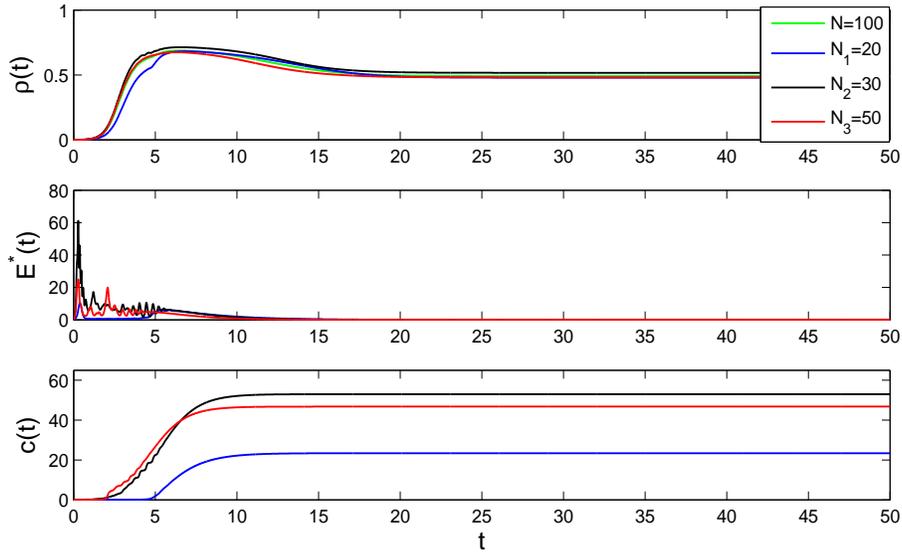}
\caption{The changes of epidemic prevalence $\rho(t)$, synchronization error $E^*(t)$, and coupling strength $c(t)$ in model (\ref{e02}) with epidemic rate $\lambda=0.6$. The network is composed of three randomly connected BA scale-free networks with size $N_1=20, N_2=30, N_3=50$. The steady-state values of epidemic prevalence are larger. Line colors are similar to that shown in Figure 2.}\label{fig4}
\end{center}
\end{figure}

From Fig.~\ref{fig3}, we can see that $\rho(t)$ of the total network and three different communities all reach a peak, then converge to similar positive numbers, and $E^*(t)$ of three communities have a small fluctuation, then converge to zero, which implies that the epidemic dynamics can successfully induce the synchronization of individuals with larger epidemic rate $\lambda=0.4>\lambda_c=\frac{1}{\alpha\rho(A)}$. Moreover, the epidemic prevalence $\rho(t)$ of the total network towards the steady-value, the epidemic prevalence of three communities reach the corresponding steady-values, and the three coupling strengths $c(t)$ also reach the steady-state values. In Fig.~\ref{fig4}, we increase the the epidemic rate to $0.6$, we find the epidemic dynamics enhance the speed of synchronization, and $\rho(t)$ towards the higher steady-values. The changes of the other variables are similar to Fig.~\ref{fig3}.

\begin{figure}[]
\begin{center}
\includegraphics[width=14cm,height=8cm]{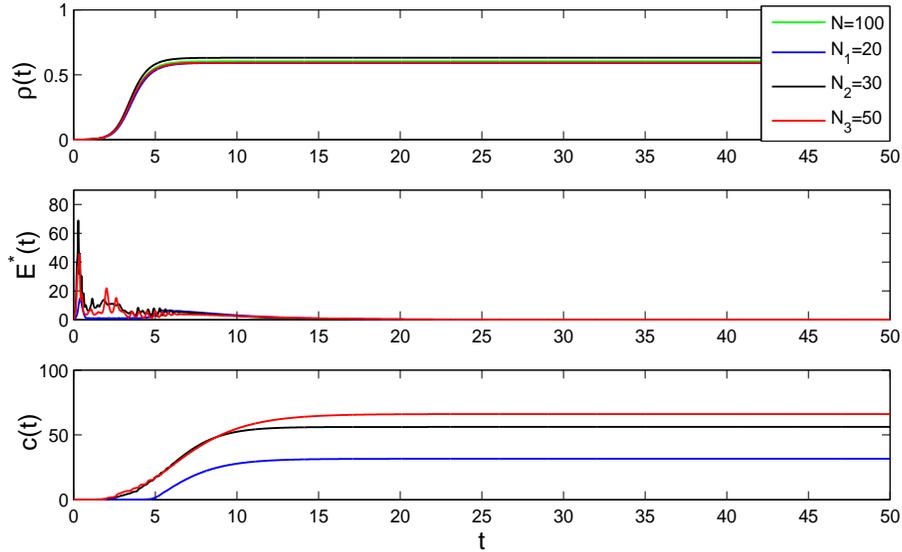}
\caption{The changes of epidemic prevalence $\rho(t)$, synchronization error $E^*(t)$, and coupling strength $c(t)$ in model (\ref{e02}) with epidemic rate $\lambda=0.4$ and $\alpha=1$. The network is composed of three randomly connected BA scale-free networks with size $N_1=20, N_2=30, N_3=50$. Line colors are similar to that shown in Figure 2.}\label{fig5}
\end{center}
\end{figure}

In Fig.~\ref{fig5}, we set $\lambda=0.4$ and $\alpha=1$, which means there is no awareness to the information of synchronization, and other variables are similar to that in Fig.~\ref{fig3}. All the epidemic prevalence $\rho(t)$ directly converge to the corresponding steady-values and $E^*(t)$ converge to zero, which implies that the epidemic dynamics can induce the synchronization of individuals with $\lambda>\lambda_c$. All the $c(t)$ of three communities are reaching to different steady states. From simulations, we can conclude that the dynamical behavior network can successfully induce the synchronization when the epidemic spreads in the network. One can see that the numerical examples illustrate these theoretical results very well.

\newpage

\begin{figure}[]
\begin{center}
\includegraphics[width=14cm,height=8cm]{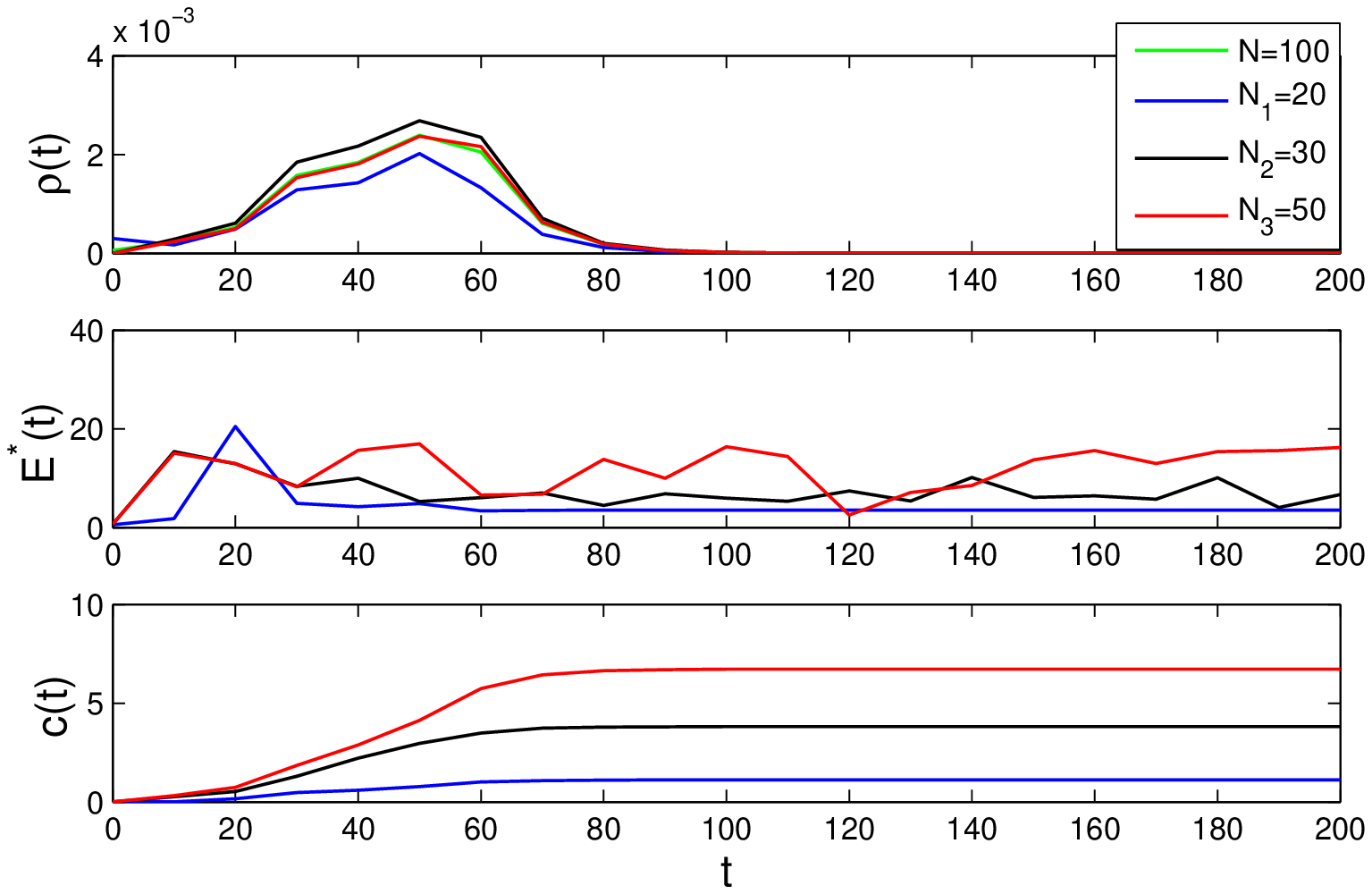}
\caption{The changes of epidemic prevalence $\rho(t)$, synchronization error $E^*(t)$, and coupling strength $c(t)$ in model (\ref{e02}) with epidemic rate $\lambda=0.145$ and the smaller initial value $\rho_3(0)=0.003$.}\label{fig6}
\end{center}
\end{figure}

\begin{figure}[]
\begin{center}
\includegraphics[width=14cm,height=8cm]{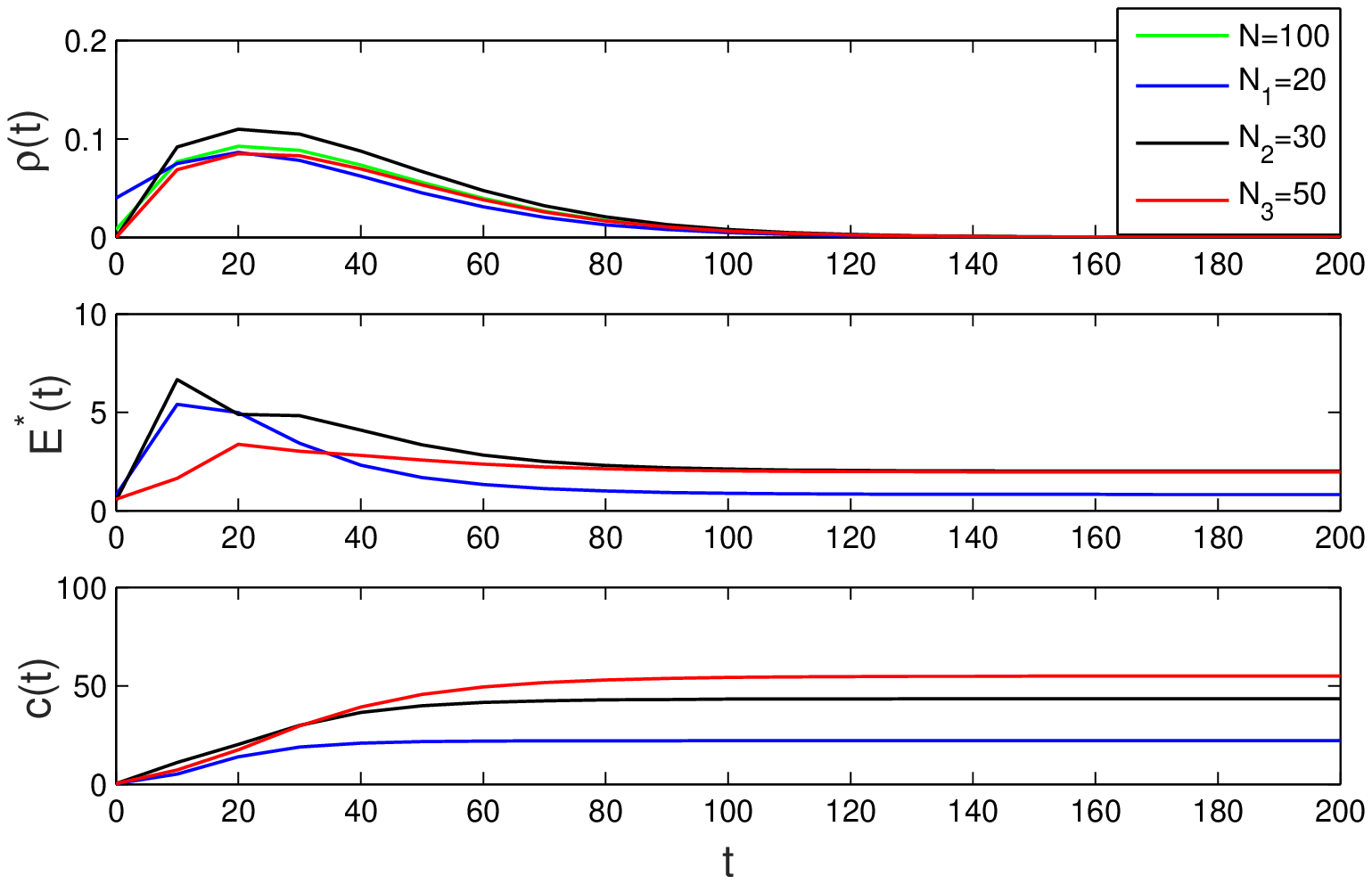}
\caption{The changes of epidemic prevalence $\rho(t)$, synchronization error $E^*(t)$, and coupling strength $c(t)$ in model (\ref{e02}) with epidemic rate $\lambda=0.155$ and the larger initial value $\rho_3(0)=0.35$. }\label{fig7}
\end{center}
\end{figure}

\begin{figure}[]
\begin{center}
\includegraphics[width=14cm,height=8cm]{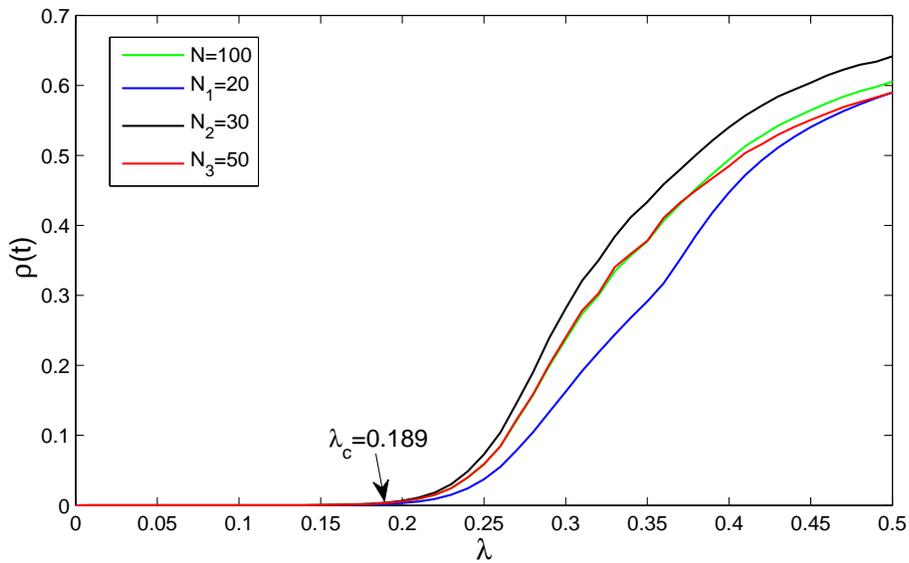}
\caption{The epidemic prevalence $\rho(t)$ of the total network and three different communities respect to the effective spreading rate $\lambda$, respectively.}\label{fig8}
\end{center}
\end{figure}

\noindent\textbf{Remark}
    \emph{In the theoretical part, there exist two critical values $\frac{1}{\rho(A)}$ and $\frac{1}{\alpha\rho(A)}$. Through some simulations, one can note that there is no backward bifurcation for the model when the effective spreading rate changes between these two critical values. In Fig.~\ref{fig6}-\ref{fig7}, we set $\lambda=0.145$ or $0.155\in(\frac{1}{\rho(A)},\ \frac{1}{\alpha\rho(A)})$ with different initial values, respectively. In these cases, the epidemic prevalence of the network converge to zero, which denote the disease will die out. The synchronization errors in three communities do not coverage to zero, that is, the epidemic dynamics cannot successfully induce the synchronization of individuals. Fig.~\ref{fig8} shows the relationship between the effective spreading rate $\lambda$ and the epidemic prevalence of the total network and three different communities. When $\lambda$ reaches the critical value $\lambda_c=0.189$, the plots all begin to rise, i.e., the disease could begin to spread.}

\section{Conclusions}

On account of the synchronization behavior of individuals induced by epidemic dynamics in real epidemic networks, this paper has established a mathematical model of dynamical networks with community structure and epidemic spreading network. We consider the cluster synchronization of community networks with nonidentical nodes, and the identical local dynamics for all individual nodes in each community. The standard SIS epidemic spreading model in a quenched
network is considered. The epidemic threshold of spreading network (\ref{eq02}) is obtained by applying $Ger\breve{s}gorin$ Lemma and dynamical system theory, and it depends on constant $\alpha$ and the spectral radius of the adjacency matrix. By Lyapunov function method and numerical simulations, it is show that the epidemic threshold may equal to $\frac{1}{\alpha\rho(A)}$, i.e., the disease-free equilibrium of the system (\ref{e02}) is globally stable in $H$ while $\lambda<\frac{1}{\alpha\rho(A)}$, otherwise, $\rho_0$ is unstable and the system is uniformly persistent in $H^0$ if $\lambda>\frac{1}{\alpha\rho(A)}$. By Lyapunov stability theory, we study the correlations between the stability of cluster synchronization and the spreading rate of the model, the conclusion is that the synchronization manifold of the dynamical behavior is globally asymptotically stable if the effective transmission rate $\lambda>\lambda_c$, i.e., the infection becomes endemic. Hence, this work provides a new perspective to the study between cluster synchronization and epidemic spread in complex networks.

The simulations show that if the disease dies out, the epidemic prevalence $\rho(t)$ converges to zero, while the synchronization error $E^*(t)$ does not coverage to zero, which means that the epidemic dynamics cannot induce the synchronization successfully. On the contrary, we can see that $\rho(t)$ converges to the steady-value, and $E^*(t)$ converges to zero, which implies that the epidemic dynamics can successfully induce the cluster synchronization of individual behaviors when the disease is persistent. When the effective spreading rate between these two critical values, the disease will also die out. Our work is the first one to address cluster synchronization in community networks with nonidentical nodes and epidemic transmission, so we hope our results provide a new insight into studying synchronization and epidemic spread in complex networks.

\section*{Acknowledgments}
This work was jointly supported by the NSFC grants under Grant Nos. 11572181, 11331009 and 61663006.
And we thank Wei Zhang from Beijing Jiaotong University for his kind help in numerical simulations.


\begin{thebibliography}{99}

\bibitem{1} \newblock G. Orosz, R.E. Wilson, G. St$\acute{e}$p$\acute{a}$n, \newblock Traffic jams: dynamics and control, \newblock \emph{Phil. Trans. Royal Society A}, \textbf{368}(1928) (2010), 4455-4479.

\bibitem{2} \newblock M.N. Dharmaweera, R. Parthiban, Y.A. Sekercioglu, \newblock Toward a power-efficient backbone network: The state of research, \newblock \emph{IEEE Commun. Surv. Tut.}, \textbf{17}(1) (2015), 198-227.

\bibitem{3} \newblock M. Vidal, M.E. Cusick, A.L. Barab$\acute{a}$si, \newblock Interactome networks and human disease, \newblock \emph{Cell}, \textbf{144}(6) (2011), 986-995.

\bibitem{4} \newblock Z.U. Khan, M. Hayat, M.A. Khan, \newblock Discrimination of acidic and alkaline enzyme using Chou's pseudo amino acid composition in conjunction with probabilistic neural network model, \newblock \emph{J. Theor. Biol.}, \textbf{365c}(6) (2015), 197-203.

\bibitem{5} \newblock S. Vitali, J. Glattfelder, S. Battiston, \newblock The network of global corporate control, \newblock \emph{Plos One}, \textbf{6}(10) (2011), e25995/1-6.

\bibitem{6} \newblock A. Garas, P. Argyrakis, C. Rozenblat, M. Tomassini, S. Havlin, \newblock Worldwide spreading of economic crisis, \newblock \emph{New J. Phys.}, \textbf{12}(2) (2010), 185-188.

\bibitem{7} \newblock D.J. Watts, \newblock A 21st century science, \newblock \emph{Nature}, \textbf{445}(7127) (2007), 489.

\bibitem{8} \newblock A.L. Barab$\acute{a}$si, \newblock Scale-free networks: a decade and beyond, \newblock \emph{Science}, \textbf{316}(5827) (2007), 1036-1039.

\bibitem{9} \newblock M.E.J. Newman,
\newblock Spread of epidemic disease on networks, \newblock \emph{Phys. Rev. E}, \textbf{66}(1) (2002), 016128/1-11.

\bibitem{10} \newblock X.C. Fu, M. Small, D.M. Walker, H.F. Zhang, \newblock Epidmeic dynamics on scale-free networks with piecewise linear infectivity and immunization, \newblock \emph{Phys. Rev. E}, \textbf{77}(3) (2008), 036113/1-8.

\bibitem{11} \newblock C.C. McCluskey, \newblock Global stability for an SIR epidemic model with delay and nonlinear incidence, \newblock \emph{Nonlinear Anal.: Real World Appl.}, \textbf{11}(4) (2010), 3106-3109.

\bibitem{12}  \newblock K.Z. Li, M. Small, H.F. Zhang, X.C. Fu, \newblock Epidemic outbreaks on networks with effective contacts, \newblock \emph{Nonlinear Anal.: Real World Appl.}, \textbf{11}(2) (2010), 1017-1025.

\bibitem{13a} \newblock M.F. Sun, Y.J. Lou, J.Q. Duan, X.C. Fu, \newblock Behavioral synchronization induced by epidemic spread in complex networks, \newblock Chaos, \textbf{27} (2017), 063101-1/14.

\bibitem{13} \newblock B. Mauricio, L.M. Pecora, \newblock Synchronization in small-world systems, \newblock \emph{Phys. Rev. Lett.}, \textbf{89}(5) (2002), 265-269.

\bibitem{14} \newblock L. Kocarev, U. Parlitz, \newblock Generalized synchronization, predictability, and equivalence of unidirectionally coupled dynamical systems, \newblock \emph{Phys. Rev. Lett.}, \textbf{76}(11) (1996), 1816-1819.

\bibitem{15} \newblock H.N. Agiza, M.T. Yassen, \newblock Synchronization of Rossler and Chen chaotic dynamical systems using active control, \newblock \emph{Phys. Lett. A}, \textbf{278}(00) (2001), 191-197.

\bibitem{16} \newblock L.M. Pecora, T.L. Carroll, \newblock Synchronization in chaotic systems, \newblock \emph{Phys. Rev. Lett.}, \textbf{64}(8) (1990), 821-824.

\bibitem{17} \newblock J.H. L\"{u}, X.H. Yu, G.R. Chen, \newblock Chaos synchronization of general complex dynamical networks, \newblock \emph{Phys. A}, \textbf{334}(1) (2004),  281-302.

\bibitem{18} \newblock C. Zhou, J. Kurths, \newblock Dynamical weights and enhanced synchronization in adaptive complex networks, \newblock \emph{Phys. Rev. Lett.}, \textbf{96}(16) (2006), 164102/1-4.

\bibitem{19} \newblock S.H. Strogatz, I. Stewart, \newblock Coupled oscillators and biological synchronization, \newblock \emph{Sci. Am.}, \textbf{269} (1993), 102-109.

\bibitem{20} \newblock C.M. Gray, \newblock Synchronous oscillations in neuronal systems: mechanisms and functions, \newblock \emph{Comput J. Neurosci.}, \textbf{1}(1) (1994), 11-38.

\bibitem{21} \newblock Z. N$\acute{e}$da, E. Ravasz, T. Vicsek, Y. Brechet, A.L. barab$\acute{a}$si, \newblock Physics of the rhythmic applause, \newblock \emph{Phys. Rev. E}, \textbf{61}(6) (2000), 6987-6992.

\bibitem{22} \newblock K.Z. Li, X.C. Fu, M. Small, Z.J. Ma, \newblock Adaptive mechanism between dynamical synchronization and epidemic behavior on complex networks, \newblock \emph{Chaos}, \textbf{21}(3), (2011), 033111/1-6.

\bibitem{23} \newblock K.Z. Li, Z.J. Ma, Z. Jia, M. Small, X.C. Fu, \newblock Interplay between collective behavior and spreading dynamics on complex networks, \newblock \emph{Chaos}, \textbf{22}(4), (2012), 043113/1-10.

\bibitem{24} \newblock M.E.J. Newman, \newblock The structure and function of complex networks, \newblock \emph{SIAM Rev.}, \textbf{45}(2), (2003), 167-256.

\bibitem{25} \newblock M.E.J. Newman, M. Girvan, \newblock Finding and evaluating community structure in networks, \newblock \emph{Phys. Rev. E}, \textbf{69}(2), (2004), 026113/1-15.

\bibitem{26} \newblock K.H. Wang, X.C. Fu, K.Z. Li, \newblock Cluster synchronization in community networks with nonidentical nodes, \newblock \emph{Chaos}, \textbf{19}(2), (2009), 023106/1-10.

\bibitem{27} \newblock Q.C. Wu, X.C. Fu, M. Small, X.J. Xu, \newblock The impact of awareness on
epidemic spreading in networks, \newblock \emph{Chaos}, \textbf{22}(1), (2012), 013101/1-8.

\bibitem{28} \newblock F.Z. Zhang, \newblock Matrix Theory: Basic results and techniques(Universitext) (New York: Springer), (2010), pp68.

\bibitem{29} \newblock H.B. Guo, M.Y. Li, Z.S. Shuai, \newblock Global stability of the endemic equilibrium of multigroup SIR epidemic models, \newblock \emph{Can. Appl. Math. Q.}, \textbf{14}(259), (2006), 259-284.

\bibitem{30} \newblock A.L. Barab$\acute{a}$si, R. Albert, \newblock Emergence of scaling in random networks, \newblock \emph{Science}, \textbf{286}(5439), (1999), 509-512.

\end{thebibliography}
\end{document}